\documentclass[12pt]{article}
\usepackage{epsfig}

\newcommand{\beq}{\begin{equation}}
\newcommand{\eeq}{\end{equation}}
\newcommand{\bea}{\begin{eqnarray}}
\newcommand{\beas}{\begin{eqnarray*}}
\newcommand{\beau}[1]{\begin{equation} \label{#1} \begin{array}{rcl}}
\newcommand{\eea}{\end{eqnarray}}
\newcommand{\eeas}{\end{eqnarray*}}
\newcommand{\eeau}{\end{array} \end{equation}}
\newcommand{\bay}{\begin{array}}
\newcommand{\eay}{\end{array}}

\newcommand{\vev}[1]{\langle #1 \rangle}

\newcommand{\esp}[1]{\, \mbox{\large \sl e}^{\textstyle \, #1}}

\begin{document}


\begin{titlepage}

\setlength{\textheight}{23.5cm}

\begin{center}
{\Large \bf Hard Parton Rescatterings and Minijets \\
in Nuclear Collisions at the LHC} \vspace{1cm} \\
        { \bf Alberto Accardi}\footnote{E-mail:{\it accardi@ts.infn.it}}
    and
    { \bf Daniele Treleani}\footnote{E-mail:{\it daniel@ts.infn.it}}
\vspace*{.5cm}  \\
        {\it Dipartimento di Fisica Teorica, Universit\`a di Trieste, \\
        Strada Costiera 11, I-34014 Trieste}  \\
    and \\
    {\it INFN, Sezione di Trieste  \\
        via Valerio 2, I-34127 Trieste }
\vspace{.4cm} \\
\end{center}

\vspace{1cm}
\begin{abstract}
The average number of minijets and the corresponding transverse
energy produced in heavy ion collisions are evaluated by including
explicitly semi-hard parton rescatterings in the dynamics of the
interaction. At the LHC semi-hard rescatterings have a sizable
effect on global characteristics of the typical inelastic event.
An interesting feature is that the dependence on the cutoff which
separates soft and hard parton interactions becomes less critical
after taking rescatterings into account.
\end{abstract}

\begin{flushbottom}
\begin{footnotesize}
\centerline{PACS: 11.80.La, 24.85.+p, 25.75.-q}
\end{footnotesize}
\end{flushbottom}

\end{titlepage}

\setcounter{page}{2}
\setcounter{footnote}{0}


\section{Introduction}
\setcounter{equation}{0}

Semi-hard physics is one of the most important issues in the
interaction of heavy ions at the LHC. Given the total energy
involved and the large number of participants, the component of
the inelastic interaction which can be described within a
perturbative approach is in fact rather substantial in heavy ion
collisions at the LHC \cite{mjgen,K00}. The result is the production of
a large number of minijets in the typical inelastic event. The
description of the semi-hard component of the interaction, adopted
by the majority of the papers on the subject, follows the approach
of ref.\cite{Kajantie:1987pd}: the semi-hard component of the
inelastic cross section, in a collision of two nuclei with atomic
mass numbers $A$ and $B$, is written as: \bea
    \sigma_H &=& \int d^2\beta
        \left( 1 - \esp{-\sigma_JT_{AB}(\beta)} \right)
        \nonumber \\
    &=& \sum_{n=1}^{\infty}\int d^2\beta \,
        {\left(\sigma_JT_{AB}(\beta)\right)^n \over n!}
    \esp{-\sigma_J T_{AB}(\beta)} \ ,
    \label{Poisson}
\eea
where
\bea
    \sigma_J&=&\int_{xx's>4p_0^2}d xd x'
        \sum_{ff^\prime}G_A^{f}(x)
        \sigma^{ff^\prime}(xx')
        G_B^{f^\prime}(x')
        \nonumber \\
    T_{AB}(\beta)&=&\int d^2b \, \tau_A(b)\tau_B(b-\beta) \ .
 \label{Def1}
\eea
In Eq.(\ref{Def1}) $\tau_A(b)$ and $\tau_{B}(b-\beta)$ are
the thickness functions of the two interacting nuclei, normalized
to A and B, respectively, and depending on the transverse coordinates of the
interacting partons $b$ and $b-\beta$, where $\beta$ is the impact
parameter of the nuclear collision. $\sigma_J$ is the single
scattering cross section to produce jets, expressed as a
convolution of the nuclear parton distributions $G_A^{f}(x)$,
$G_B^{f^\prime}(x')$ and of the partonic cross section
$\sigma^{ff^\prime}$ (integrated on the momentum transfer with the cutoff $p_0$
and without the
multiplicity factor of the produced jets). The indices $f$ and $f^\prime$ label the
different kinds of interacting partons and the momentum fractions
$x$, $x'$ are defined with respect to the single nucleon momentum.
The dependence on the scale factor is
implicit in all quantities.

The semi-hard cross section, as expressed in Eq.(\ref{Poisson}),
represents the contribution to the total inelastic cross section
of all events with at least one semi-hard partonic interaction, so
that one may write $\sigma_{inel}=\sigma_H+\sigma_{soft}$, the two
components being separated by means of $p_0$. Eq.(\ref{Poisson})
may be derived in an eikonal approach to nucleus-nucleus
interactions, by writing the eikonal phase as the sum of a soft
and of a hard component \cite{mjgen,Capella:1987cm}. The physical
picture corresponding to Eq.(\ref{Poisson}) is that of a
distribution of multiple independent parton collisions localized
in different points in transverse space and with the average
number depending on the nuclear impact parameter. The average
number of parton interactions, at fixed $\beta$, is $\langle
N(\beta)\rangle=\sigma_J T_{AB}(\beta)$ so that, if $\sigma_J$ is
at the lowest order in $\alpha_S$, the average multiplicity of
minijets in a given nuclear collision (that is at $\beta$ fixed)
is $2\langle N(\beta)\rangle$. The integrated inclusive cross
section for producing minijets is therefore given by $2\int
d^2\beta\langle N(\beta)\rangle$. To obtain the inclusive cross
section one needs, in fact, to count the jets produced with their
multiplicity, so that, in the case of multiple parton
interactions, the inclusive cross section is obtained by taking
the average of the distribution in the number of parton
collisions. The result is that the integrated inclusive cross
section is given by the single scattering expression, $\sigma_J$,
multiplied by the average multiplicity of jets produced in a
single partonic interaction, which shows that the description of
the process, as given by Eq.(\ref{Poisson}), is consistent with
the AGK cancellation \cite{agk} (all unitarity corrections cancel
in the inclusive). The cancellation property obviously holds for
all averages, that are therefore equal to the result obtained by
means of the single scattering  expression, so that the transverse
energy produced by minijets is given by \beq
    \langle E_t(\beta)\rangle=2\, T_{AB}(\beta)\int_{p_t\ge p_0}
        p_t\frac{d\sigma_J}{d^2p_t}\, d^2p_t \ .
  \label{Et}
\eeq
The semi-hard cross section $\sigma_H$ is a smooth function of $p_0$ for
small values of the cut-off: the limiting value of $\sigma_H$ is in fact the
geometrical limit $\pi(R_A+R_B)^2$, $R_A$ and $R_B$ being the two
nuclear radii. On the contrary $\langle N(\beta)\rangle$ and
$\langle E_t(\beta)\rangle$ are singular at small $p_0$ and their
behaviour may be roughly estimated on dimensional grounds to be $\langle
N(\beta)\rangle\simeq1/p_0^2$ and $\langle
E_t(\beta)\rangle\simeq1/p_0$. The singular behavior at low $p_0$
can be used to set the limits of validity of the picture. Indeed, for the
picture of the interaction to be valid one should take a relatively large
value of the cut-off $p_0$; in this way the whole semi-hard interaction takes
place in a relatively dilute system and the overall number of interactions
will be relatively small. To
deal with a regime where the number of parton interactions
and the density of the interacting partons are large, the main
modification adopted by the majority of papers is
to include shadowing corrections in the nuclear parton
distributions \cite{shad,EKKV00}. In this way one obtains a substantial
reduction of the number of projectile and target partons at low
$x$ and the picture can be extended to sizably lower values of the
cutoff $p_0$. Even so, when $p_0$ is further reduced, one reaches
the condition of a highly dense interacting system where the whole
picture ceases to be valid, which sets the lower limit for a
sensible choice of the cutoff $p_0$ \cite{GLR83,satur}. \\
The
overall resulting features are therefore that $p_0$ is different
when varying the atomic mass number of the interacting nuclei and
their energy, and that the distribution in the number of hard
collisions at a fixed value of the impact parameter $\beta$ is a
Poissonian in the whole semi-hard regime so that all
average quantities are computed, as above, with the single
scattering  expression.

The clean physical interpretation of the approach, which
incorporates the geometrical features of the nuclear process,
unitarity, the factorization of the hard component of the
interaction and the AGK cancellation rule, justifies the great
success of the picture. Still there are a few delicate points
which deserve further investigation and where the description of
the process might be improved.

\begin{figure}[t]
\begin{center}
\epsfig{figure=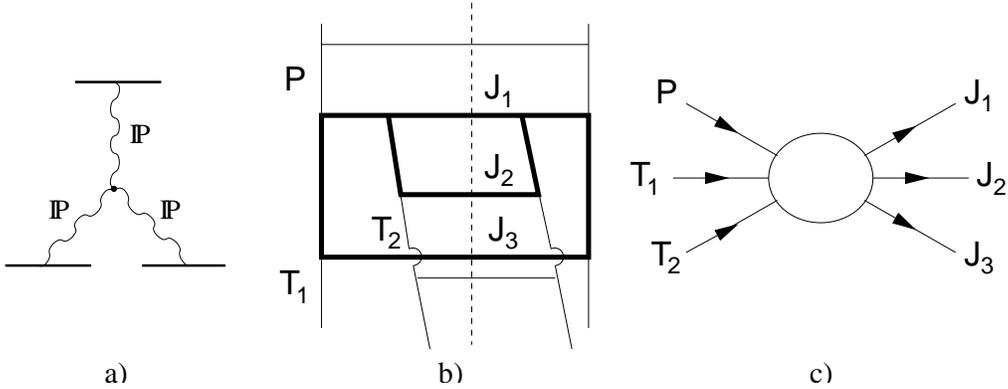,height=2in} \caption{\footnotesize
a) A triple pomeron interaction between 3 nucleons. b) Parton ladders with the
topology of Fig.\ref{regge}a). The thick lines are fluctuations with high
$p_t$ and the dashed line is a possible cut. c) Partonic view of the cut of 
Fig.\ref{regge}b): P is a projectile parton interacting with two target
partons T$_i$ and the final state is given by the three large-$p_t$ jets
J$_i$.
  \label{regge}  }
\end{center}
\end{figure}

A very general approach to nucleus-nucleus interactions is trough
the Reggeon Diagram Technique, where at high energies the interaction is
described by the exchange of many Pomerons, including both 
independent exchanges between different nucleons and multi-Pomeron
interactions, which represent the collision of a given projectile
nucleon with several different target nucleons in a given
interaction process. Each partonic collision corresponds to a
fluctuation with a large transverse momentum inside the structure
of an exchanged Pomeron. While the simplest case is that with a
single partonic loop with large $p_t$, whose discontinuity
corresponds to a $2\to2$ partonic collision, one might imagine
more complicated fluctuations, with several connected parton
lines, all with large $p_t$, in the structure of a multi-Pomeron
interaction. The simplest case of this kind is illustrated in
Fig.\ref{regge}, where the Pomerons are represented by parton ladders.
Discontinuities of such fluctuations originate configurations,
where the compensation of transverse momenta in the final state
involves several large $p_t$ partons, while the whole large $p_t$
configuration is generated by partons belonging to different
chains, representing Pomerons attached to different nucleon lines.
In the picture of the semi-hard interaction just recalled those
configurations contribute to the shadowing corrections of the
nuclear parton distributions. Their discontinuities, on the
contrary, are not included in the semi-hard interaction dynamics.
On the other hand, when approaching the black disk limit, the
initial and final partonic states become locally isotropic in
transverse space, so that both initial and final state
configurations need to be built up with lots of partons. A proper
discussion of semi-hard dynamics in the black disk limit requires
therefore taking into account fluctuations with many large $p_t$
parton lines, which are to be included not only in the virtual
corrections but also in the production process. Hence one needs to
consider partonic interactions where several partons, with low
virtuality and $p_t$ and sizable longitudinal components (so that,
in the nucleus-nucleus c.m. frame, each one may be ascribed to a
definite nucleus) interact producing large $p_t$ jets. The
simplest possibility of this kind (one projectile against two
targets, $t/s\to 0$ and basically electro-dynamical interaction
between point like objects) was discussed in Ref.\cite{eik}, where
the forward amplitude of the process and all the cuts were
derived. The final result of the analysis was that, in the $t/s\to
0$ limit, the different cuts of the amplitude are all proportional
one to another, the proportionality factors being the AGK weights.
It was moreover possible to express the three body interaction as
a product of two body  interaction probabilities. Therefore, one may
argue that, in the black disk limit and in the
nucleus-nucleus c.m. frame, a projectile parton interacts with all
the various nearby partons of the target in the different
directions in transverse space, and that the corresponding cross
section may be approximated by the product of two-body interaction
cross sections, so that one might call such a process {\it re-interaction}
or {\it re-scattering}.

An attempt to introduce such more elaborate semi-hard dynamics,
including explicitly semi-hard parton rescatterings in the
interaction, was done in ref. \cite{ct1} and \cite{ct2}. Both the
average number of minijets and the average transverse energy are
modified by semi-hard rescatterings, and an interesting feature is
that both quantities develop a less singular dependence on the
cutoff, in such a way that the choice of $p_0$ becomes less
critical when semi-hard parton rescatterings are taken into
account. The average number of minijets and the transverse energy
produced in heavy ion collisions have been recently discussed in
several papers, with the purpose to determine the initial
conditions for the further evolution and termalization of the
system (see e.g. \cite{EKKV00,EKR98}). We think that it might be
interesting to have an indication of the effects of rescatterings
on these quantities at the LHC, and we'll discuss this topic in
the present paper. In the next section, after including
rescatterings in the picture of the interaction, we'll recall the
expression of the average number of minijets and derive the
corresponding average transverse energy. Then, we'll give some
quantitative indication on the effect and comment the qualitative
features induced by the more structured interaction dynamics.


\section{Multiple Parton Scatterings and Average Number of Minijets}
\setcounter{equation}{0}

The introduction of semi-hard parton rescatterings can be obtained in a
picture of the interaction where the soft part is factorized in a
Poissonian multi-parton distribution and the hard part is
expressed in terms of perturbative parton-parton collisions. The nuclear
multi-parton distributions are unknown quantities and the reason
of the choice of the Poisson distribution is that it corresponds
to the case where the information on the initial state is minimal,
since the whole distribution is expressed in terms of its average
value only. One has moreover the possibility of introducing
systematically further informations on the nuclear partonic
structure in terms of correlations among partons \cite{ctresc}.
The hard part is written in terms of two-body
collisions by introducing the probability of having at least one
interaction between the two configuration, in a way analogous to
the expression of the inelastic nucleus-nucleus cross
section \cite{bbc}. The process is
therefore represented as the sum of all possible interactions
between all configurations with a definite number of partons of
the two nuclei. In this way the semi-hard component of the cross
section can be written as

\bea
    \sigma_H &=&\int d^2\beta \, \sum_{n=1}^{\infty} \, {1\over n!}
        \, \Gamma_A(x_1,b_1-\beta) \dots
        \Gamma_A(x_n,b_n-\beta) \,
        \esp{-\int d x d^2b \, \Gamma_A(x,b-\beta)} \cdot
        \nonumber \\
    && \hspace*{1.1cm} \cdot \sum_{l=1}^{\infty}{1\over l!} \,
        \Gamma_B(x_1',{b}_1') \dots
        \Gamma_B(x_l',{b}_l') \,
        \esp{{-\int d x' d^2b' \Gamma_B(x',{b}')}} \cdot
        \nonumber \\
    && \hspace{1.1cm} \cdot \, \Bigl[1-\prod_{i=1}^n
        \prod_{j=1}^l(1-\hat{\sigma}_{ij})\Bigr] \,
         dx_1d^2b_1\dots dx_nd^2b_n
        dx_1'd^2b_1'\dots dx_l'd^2b_l' \ \, ,
  \label{csect}
\eea
where
\bea
    \Gamma_A(x,b)=\tau_A(b)G(x) \ . \nonumber
\eea
To keep the notation as simple as possible, the
indices labeling the different kinds of partons have been
suppressed and the dependence on the cutoff $p_0$ is implicit. The
two Poissonian distributions, with average numbers $\Gamma_A(x,b-\beta)$ and
$\Gamma_B(x',b')$, represent the multi-parton
distributions of the two interacting nuclei. The probability to
have at least one partonic interaction, given a configuration with
$n$ and $l$ partons in the two nuclei, is represented by the
square parenthesis in Eq.(\ref{csect}) and, according with the discussion in
the previous paragraph, is constructed by means
of the probability $\hat{\sigma}_{ij}$ of interaction of a given
pair of partons $i$ and $j$. The interaction probability is a function of the
cutoff $p_0$, so that only the interactions with momentum exchange larger than
$p_0$ are taken into account in (\ref{csect}). The majority of interactions
takes place with a momentum exchange close to the cutoff value, we therefore
evaluate the parton distributions at the scale of the cutoff.
Since the distance over which the hard interactions are localized is much
smaller than the typical nuclear radius, the probability of interaction can
be approximated by $\hat{\sigma}(x_ix_j,b_i-b_j)=\sigma(x_ix_j)
\delta^{(2)}(b_i-b_j)$. \\
The most important features of Eq.(\ref{csect}) are that all possible
interactions between the two partonic configurations are included,
and that  probability conservation is explicitly taken into account
by the term in square parentheses.
The Eikonal cross-section in Eq.(\ref{Poisson}) is obtained when one neglects
all rescatterings in Eq.(\ref{csect}), in such a way that each
parton is allowed to interact at most once \cite{AT88}. If, on
the contrary, one keeps rescatterings into account one cannot
write a closed expression for $\sigma_H$. However, it is possible
to obtain simple expressions from Eq.(\ref{csect}) for many relevant
quantities. \\
If one works out the average number of parton
collisions $\langle N(\beta)\rangle$ one obtains, as in the
traditional approach, the single scattering  expression
$\langle N(\beta)\rangle=\sigma_J T_{AB}(\beta)$ \cite{ct1}. So
the overall average number of parton collisions satisfies the
AGK cancellation and is not affected by any unitarity correction; however
it is not a simple quantity to observe.
A more directly observable quantity, or at least one which
can be more directly related to observable quantities, is the
average number of produced minijets. An important effect of
including rescatterings is that the number of produced minijets is
no more proportional to the number of collisions, because now each
projectile parton is allowed to interact more than once with the
target. As a consequence, while unitarity corrections do not
change the average number of collisions, they affect the average
number of minijets produced in the nuclear collision.

The correction term can be derived in a straightforward way from
Eq.(\ref{csect}) \cite{ct1}, but one may use also a more direct argument.
One may in fact obtain the usual semi-hard cross section, Eq.(\ref{Poisson}),
by starting from the single scattering cross section to produce large-$p_t$
jets. In the perturbative QCD-parton model the cross section is written as:
\beq
    \sigma_J = \int_{xx's>4p_0} d^2\beta \, d^2b \, d x \, d x' \,
        \Gamma_A(x,b-\beta) \, \sigma(xx') \, \Gamma_B(x',b)
        = \int d^2\beta \, \langle N(\beta) \rangle \ .
  \label{pqcd}
\eeq
The expression needs however to be unitarized also when the cutoff $p_0$ has a rather
large value, since $\sigma_J$ is proportional
to the large factor $A\times B$.
Eq.(\ref{Poisson}) is in fact the result of the $s$-channel
unitarization of $\sigma_J$ and may be obtained by noticing that
in Eq.(\ref{pqcd}) $\langle
N(\beta)\rangle$ is dimensionless and may be understood as
the average number of parton interactions at a given impact parameter
$\beta$. The $s$-channel unitarized cross section
$\sigma_H$ is the result of replacing this average number
with the interaction probability, which, if the
distribution in the number of interactions is a Poissonian, is
just $1-\exp(-\langle N(\beta)\rangle)$. Hence the unitarized
cross section $\sigma_H$ represents the contribution to the total
cross section of all events with at least one couple of partons
interacting with a transverse momentum exchange above $p_0$, as it
is clear from the second line of Eq.(\ref{Poisson}). On the other hand
$\sigma_J$, that includes also the multiplicity of the
interactions, represents rather the integrated inclusive cross
section (apart from the factor representing the average multiplicity
of jets produced in a single collision). \\
When the cutoff is moved towards low values and rescatterings need
to be taken into account, the average number of jets produced is
no more proportional to the average number of collisions. In this
case one may proceed by applying to $\langle N(\beta)\rangle$
an argument analogous to that previously used to unitarize $\sigma_J$.
By looking at Eq.(\ref{pqcd}) one can identify
\bea
    \vev{n_B(x,b)} \equiv \int_{xx's>4p_0^2} d x' \,
        \Gamma_B(x',b) \,  \sigma(xx')
  \label{nb}
\eea
as the average number of collisions of each interacting A-parton at fixed $x$
and $b$. Then one can write the average number of produced minijets at fixed
impact parameter as
\beq
        2\langle N(\beta)\rangle = \int_{xs>4p_0^2} d^2bdx \Gamma_A(x,b-\beta)
            \, \langle n_B(x,b)\rangle \nonumber
        + \int_{x's>4p_0^2} d^2bdx' \Gamma_B(x',b)
        \, \langle n_A(x',b-\beta) \rangle \ ,
  \label{sn}
\eeq
that represents the average number of incoming partons from the nucleus A
multiplied by the average number of collisions against the partons of nucleus
B, plus  the analogous term with A and B interchanged. Then, if one replaces
in Eq.(\ref{sn}) the average number of scatterings of each parton
with its interaction probability, viz. $1-\exp(\langle
n_B(x,b)\rangle)$ and $1-\exp(\langle
n_A(x,b-\beta)\rangle)$, one obtains the average number
of ``wounded partons'' $W_A(x,b)$ and $W_B(x',b-\beta)$
of the two nuclei. These ones are the partons of the two nuclei that had
at least one hard interaction. The expression of the average number
of wounded partons of nucleus $A$ (with transverse coordinate $b$ and
momentum fraction $x$, in an event with nuclear impact parameter $\beta$)
is therefore
\beq
     W_A(x,b,\beta)=\Gamma_A(x,b-\beta) \Bigl[1-\esp{-\langle
        n_B(x,b)\rangle}\Bigl] \ .
    \label{W}
\eeq Every wounded parton obtained in this way produces a minijet
in the final state and the transverse energy produced by semi-hard
interactions is the transverse energy carried by the wounded
partons. As a consequence both the average number of minijets and
their average transverse energy are quantities affected by the
presence of rescatterings and the corresponding correction term is
more and more important when the average number of scatterings
$\langle n_B \rangle$ becomes larger and larger, namely at low
values of the cutoff $p_0$ and (or) for large atomic mass numbers. \\
The overall number of produced minijets, i.e. the sum of the
wounded partons of nucleus $A$ with those of nucleus $B$,
obviously coincides with the usual result $2\langle
N(\beta)\rangle$, when the average number of rescatterings is
small. When the number of rescatterings is large the two
quantities are however different and, while the average number of
collisions $\langle N(\beta)\rangle$ may be divergent in the
saturation limit, the average number of wounded partons is on the
contrary well defined. In fact one obtains that the square
parenthesis in Eq.(\ref{W}) has 1 as a limiting value and, in this
limit, the average number of wounded partons is just the sum of
the average number of partons of the two interacting nuclei.

\begin{figure}[t]
\begin{center}
\epsfig{figure=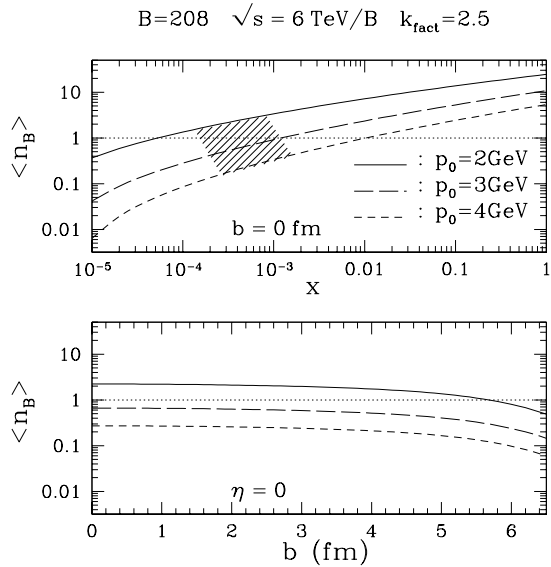,height=3in} \caption{\footnotesize
Average number $\langle n_B \rangle$ of semi-hard scatterings
suffered by a parton from the A-nucleus while it is interacting
with the B-nucleus. We consider a Pb-Pb collision at a c.m. energy
$\sqrt{s}=6$ TeV/A, with a scale $Q=p_0/2$ and a $k$-factor
$k=2.5$; Woods-Saxon thickness functions are used. {\it Top:}
$\langle n_B \rangle$ is plotted as a function of the projectile
parton fractional momentum $x$ in the case of a central
parton-nucleus collision. The three curves are the result for a
cut-off $p_0=2$ GeV (solid line), $3$ GeV (long dashes), $4$ GeV
(short dashes). The shaded area corresponds to the central
pseudo-rapidity region $|\eta|\leq.9$, where $\eta = \log
(x\sqrt(s)/p_0)$. The dotted line marks the transition value of 1
scattering per parton. {\it Bottom:} $\langle n_B \rangle$ is
plotted as a function of the impact parameter for a parton
emerging with pseudo-rapidity $\eta=0$.
  \label{nofscat}  }
\end{center}
\end{figure}

To have a quantitative feeling for the regime of interest at LHC
energies we plot $\langle n_B \rangle$, the average number of
scatterings per parton from the A-nucleus, as a function of the
projectile parton fractional momentum $x$ and impact parameter
$b$. As one can see, at a cut-off $p_0=2$ GeV a parton in the
central region experiences on average 1 to 3 scatterings over the
whole target transverse area except the very peripheral regions.
As discussed above, at higher cut-off values the number of
scatterings decreases. As a result, semi-hard rescatterings should
be negligible down to $p_0\simeq 3$ GeV where their effects begin
to show up, and already at $p_0=2$ GeV they should have a large
effect and should not be neglected. These conclusions will be
confirmed by the numerical computations discussed in section
\ref{sec:discussion}.

In summary, the effect of rescatterings on the average number of
the produced minijets is to reduce the number obtained by means of
the single scattering  expression, not differently,
qualitatively, from the
result of including shadowing corrections in the nuclear parton
distribution. On the other hand the overall distribution in the
number of minijets produced is modified. In the traditional
approach there is a strong correlation in the distribution of the
number of minijets, since each minijet has a recoiling companion; when the
average number of rescatterings increases this correlation gets
weaker and weaker, so that in the high density limit no correlation is left
and the distribution tends to a Poissonian \cite{ct1}. A further
important difference is that, since $W_A$ and $W_B$ are well
defined in this limit, after including rescatterings the average number
of minijets becomes much less dependent on the choice of the cutoff
at low $p_0$.


\section{Average Transverse Energy}
\setcounter{equation}{0}

As we have discussed in the last section, the average number of
minijets is not the only quantity modified by this more elaborate
interaction dynamics, which in fact has a non-trivial effect also
on the transverse energy carried by the minijets \cite{ct2}. The
wounded partons of nucleus $A$ at a given $x$ and $b$ and in a
nuclear interaction with impact parameter $\beta$, Eqn.(\ref{W})
are obtained by multiplying the average number of partons of $A$
(with given $x$ and $b$) by the corresponding interaction
probability, which is a Poisson probability distribution in the
number of scatterings, with average $\langle n_B(x,b)\rangle=\int
dx' \Gamma_B(x',b)  \sigma(xx')$: \beq
     W_A(x,b,\beta)=\Gamma_A(x,b-\beta)\sum_{\nu=1}^{\infty}\frac{\langle
        n_B(x,b)\rangle}{\nu!}\esp{-\langle
        n_B(x,b)\rangle} \ .
    \label{W1}
\eeq
We may therefore obtain the differential distribution in the
transverse momentum $p_t$ of the wounded $A$-partons by introducing a
constraint in the transverse momentum intergrals that give the total cross
sections in the above expression:
\bea
    \lefteqn{\frac{dW_A}{d^2p_t}(x,b,\beta) \ =} \nonumber \\
    &=& \Gamma_A(x,b-\beta)
        \sum_{\nu=1}^{\infty}\frac{1}{\nu!} \int \Gamma_B(x'_1,b)
        \dots \Gamma_B(x'_{\nu},b) \esp{ -\int dx'
        \Gamma_B(x',b)  \sigma(xx')} \, \cdot
        \nonumber \\
    && \cdot \, \frac{d\sigma}{d^2k_1} \dots \frac{d\sigma}{d^2k_{\nu}}
        \, \delta^{(2)}({\bf k}_1 + \dots + {\bf k}_{\nu}
        -{\bf p_t})
        \,\, d^2k_1 \dots d^2k_{\nu} \,\, dx_1' \dots dx_{\nu}' \ .
  \label{dWdp}
\eea The limits of integration on $x'_i$ and $x'$ are respectively
$xx'_is\geq 4k_i^2$ and $xx's\geq 4p_0^2$, and all the
distribution functions are evaluated for simplicity at a fixed
scale. To obtain the corresponding average transverse energy
$\vev{p_t(x,b,\beta)}_A$ one has to integrate Eq.(\ref{dWdp}) with
an additional factor $p_t$. A convenient way to proceed is to
introduce the Fourier transform of the parton-parton scattering
cross section \beq
    \tilde{\sigma}(u) = \int d^2k e^{i{\bf k}\cdot{\bf u}}
        \frac{d\sigma}{d^2k} \ \ \ ; \ \ \
         \tilde{\sigma}(0) = \sigma
\eeq
and to express the $\delta$-function in Eq.(\ref{dWdp}) in an integral form.
Since $\tilde{\sigma}(u)$ depends only on the modulus of ${\bf u}$ one obtains
\bea
    \langle p_t(x,b,\beta)\rangle_A &=& \Gamma_A(x,b-\beta)
        \int_0^{\infty} dp_t \, p_t^2
        \int_0^{\infty} du \, u J_0(p_tu)
        \cdot \\
    && \cdot \sum_{\nu=1}^{\infty} \frac{1}{\nu!}
        \left[\int dx' \Gamma_B(x',b)
        \tilde{\sigma}(u) \right]^{\nu}
        \esp{-\int dx' \Gamma_B(x',b)
        \tilde{\sigma}(0)} \ . \nonumber
\eea
After multiplying the integrand by the exponential factor $\exp(-\lambda p_t)$
one can exchange the two integrals and first do the integral on
$p_t$. The result of this integration is
\beq
    \int_0^{\infty} e^{-\lambda p_t} J_0(p_tu) p_t^2 dp_t
        = \frac{2\lambda^2-u^2}{(\lambda^2+u^2)^{5/2}}
        = \frac{1}{u}\frac{d}{du}
        \left[ \frac{\lambda^2u^2+u^4}{(\lambda^2+u^2)^{5/2}}
        \right] \ ,
\eeq
which allows one to perform the integral on $u$ by parts. After
differentiating the series in $\nu$ one obtains the factor
\beq
    \tilde{\sigma}'( u)=-2\pi\int_0^{\infty} k^2 J_1(ku)
        \frac{d\sigma}{d^2k} \, dk \ ,
  \label{sigp}
\eeq
which is proportional to $u$ because of the argument of the $J_1$
Bessel function. The integrand in $u$ is therefore regular for $u=0$ also in the limit
$\lambda\to0$. Then one takes this limit and sums the
series in $\nu$. The final expression has a closed analytical form:
\beq
    \vev{p_t(x,b,\beta)} = - \, \Gamma_A(x,b-\beta)
        \int_0^{\infty} du \, dx' \frac{\tilde{\sigma}'(u)}{u}
        \Gamma_B(x',b) \esp{\, \int dx' \Gamma_B(x',b)
        [\tilde{\sigma}(u)-\tilde{\sigma}(0)]}
  \label{etr}
\eeq
The average transverse energy in an event with a given impact
parameter $\beta$ is the result of integrating Eq.(\ref{etr}) on
$b$ and $x$ and of summing the two contributions of the wounded
partons of the nuclei $A$ and $B$. Notice that the expression in
Eq.(\ref{etr}) is much less dependent on the choice of the cutoff
$p_0$ than the usual average energy evaluated with the single
scattering  expression of the perturbative QCD parton model:
the Rutherford singularity of the parton-parton cross section is
in fact smoothed in Eq.(\ref{sigp}) by the Bessel function
$J_1(ku)$, in such a way that the dependence on the cutoff $p_0$
is only logarithmic. The same logarithmic dependence on the cutoff
is present in the argument of the exponential
\beq
    \tilde{\sigma}(u)-\tilde{\sigma}(0)
        =2\pi\int\Bigl[J_0(ku)-1\Bigr]\frac{d\sigma}{d^2k}
        \, k \, dk
\eeq
as a consequence of the behavior of $\bigl[J_0(ku)-1\bigr]$ for
$k\to 0$.

In summary, when the semi-hard cross section is expressed by Eq.(\ref{csect})
one may obtain, without further approximations, a closed
analytic form both for the average number of minijets and for the
corresponding average transverse energy and in both cases the
singular dependence for small values of the cutoff is smoothened by
rescatterings.


\section{Discussion}
\label{sec:discussion}
\setcounter{equation}{0}

Many papers have been recently devoted to the production of
minijets in heavy ion
collisions \cite{K00,EKKV00,satur,EKR98}. A rather
general feature is that, because of the singular behavior of the
elementary parton interaction at low momentum transfer, many
relevant quantities depend rather strongly (typically like an
inverse power) on the value of the cutoff which distinguishes soft
and hard parton interactions. The feature is unpleasant since
although one might find physical arguments to determine a
meaningful value of the cutoff \cite{GLR83,satur,Eskola:1996bp}, it is rather
difficult to fix it in a very precise way. We have therefore
tried, in the present paper, to face this issue by studying the
effect of a more elaborate interaction dynamics on the average
number of minijets produced in a nuclear collision and on
the corresponding average transverse energy.
While in the traditional picture of the
semi-hard processes each parton is allowed to interact with large
momentum transfer only once, we have included semi-hard
parton rescatterings in the dynamics of the interaction. Semi-hard
rescatterings, that are negligible when the threshold between hard
and soft processes is high, become more and more important when
the threshold is lowered and the target approaches the black disk limit.
Naively one would expect that the
inclusion of rescatterings in the picture of the interaction
might worsen the divergent behavior at low transverse momenta; on
the contrary a more careful analysis, that takes probability
conservation consistently into account, shows that the result is
just the opposite. Following \cite{ct1} and \cite{ct2} we have in
fact represented the semi-hard nuclear cross section with
Eq.(\ref{csect}), where all possible multi-parton collisions,
including rescatterings, are taken into account and the
conservation of probability is explicitly implemented. The average
number of minijets and the corresponding transverse energy, at
fixed $x$, $b$ and impact parameter $\beta$,
are then expressed in a closed analytic
form by (\ref{W}) and (\ref{etr}), whose behavior with the cutoff is
much less singular in comparison with the analogous averages
obtained without taking rescatterings into account. The reason of
this smoother behavior is that rescatterings
introduce (through $\Gamma_A$ and $\Gamma_B$) a new dimensional
quantity in (\ref{W}) and (\ref{etr}), the nuclear radius, which gives
the dimensionality to the two average quantities at small $p_0$.
When rescatterings are neglected the dimensionality at small $p_0$
is provided by the cutoff itself, and the result is that the two
quantities behave as an inverse power of the cutoff for $p_0\to0$.

Apart from the approximation of writing all connected multi-parton
processes as products of $2\to2$ partonic collisions, our approach
states on the assumption of neglecting production processes at the
partonic level (namely $2\to3$ etc. parton processes) and of using
forward kinematics in the nucleon-nucleon c.m. frame. To have a
feeling on the validity of such approximations at LHC energy, we
have evaluated the average energy of a partonic interaction in the
parton-parton center of mass frame, and the average value of momentum fraction
$x$ of a projectile parton: 
\bea
    \langle E_{c.m.}\rangle\sigma_J&=&\int d xd x'\sqrt{xx's}
        \sum_{ff^\prime}G_A^{f}(x)
        \sigma^{ff^\prime}(xx')
        G_B^{f^\prime}(x')\nonumber\\
    \langle x\rangle\sigma_J&=&\int d xd x'x
        \sum_{ff^\prime}G_A^{f}(x)
        \sigma^{ff^\prime}(xx')
        G_B^{f^\prime}(x')
\eea
When the whole rapidity range is considered typical values are $\langle
E_{c.m.}\rangle\approx25$GeV, $\langle 
x\rangle\approx3\times10^{-2}$ (corresponding to a momentum of
$\approx45$GeV, if the nucleon-nucleon c.m. energy is $6$TeV) with
$p_0=2$GeV. The average $\langle x\rangle$ becomes substantially
smaller when averaging in a narrow window in the central rapidity
region. The relatively low value of $\langle E_{c.m.}\rangle$, as
compared with the cutoff, indicates that the inclusion of $2\to3$,
or of higher order partonic processes, should not spoil the whole
approach, that could therefore represent a reasonable lowest
order approximation. The relatively large value of the momentum
boost to go from the nucleon-nucleon c.m. frame to the partonic
c.m. frame shows, on the other hand, that forward kinematics is
reasonable in the former frame of reference.

\begin{figure}[t]
\begin{center}
\epsfig{figure=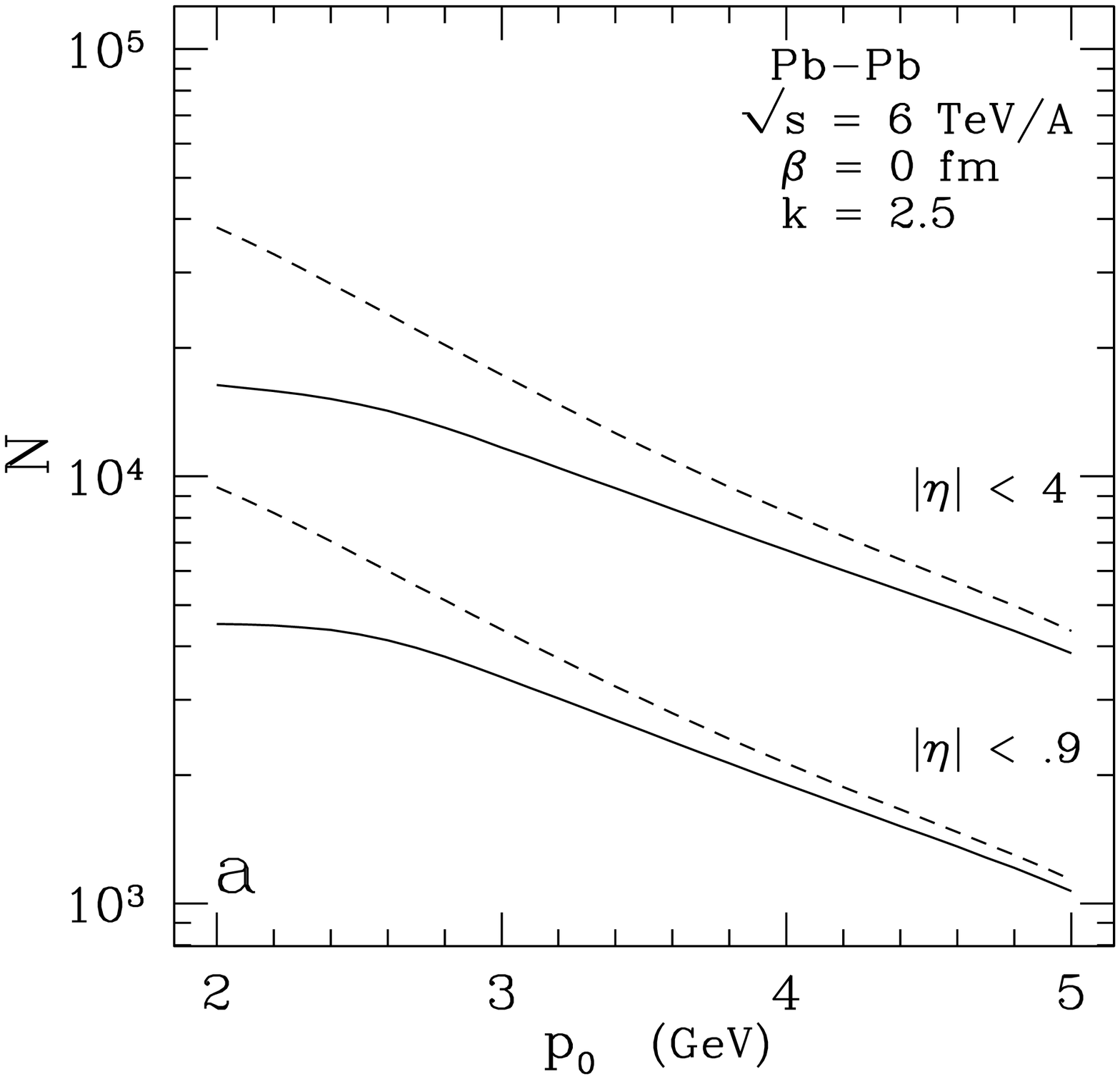,height=2.5in}
\epsfig{figure=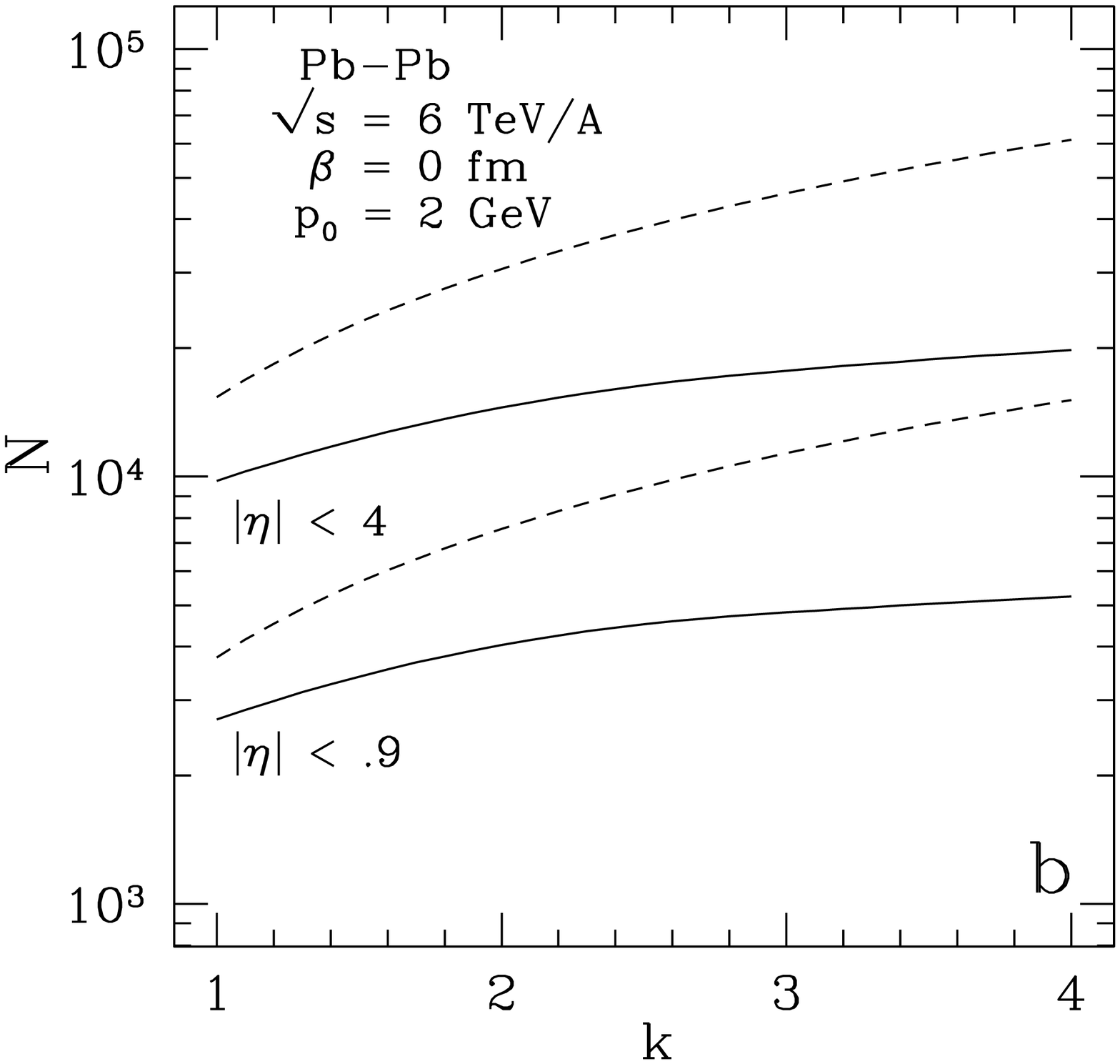,height=2.5in}
\epsfig{figure=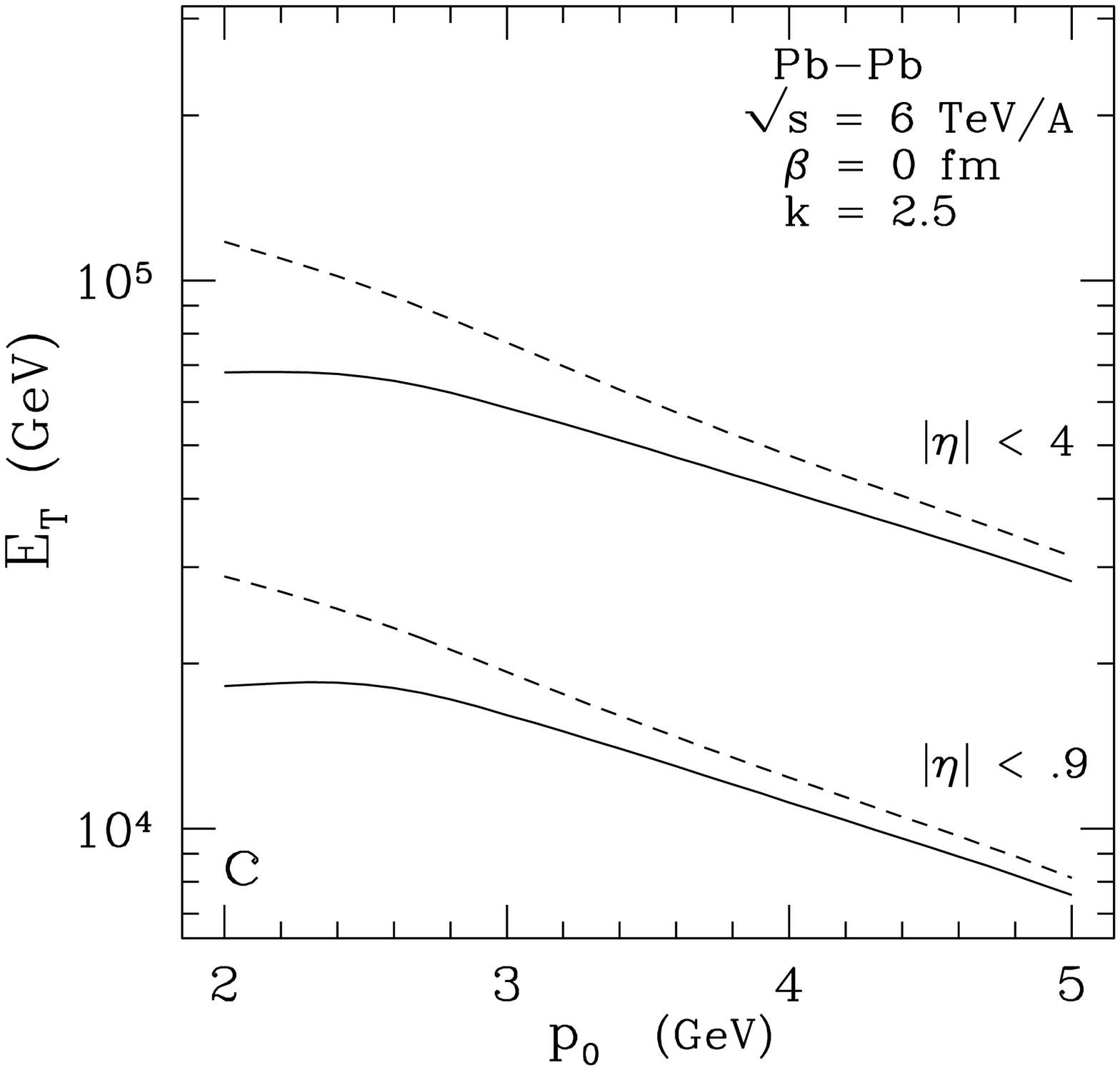,height=2.5in}
\epsfig{figure=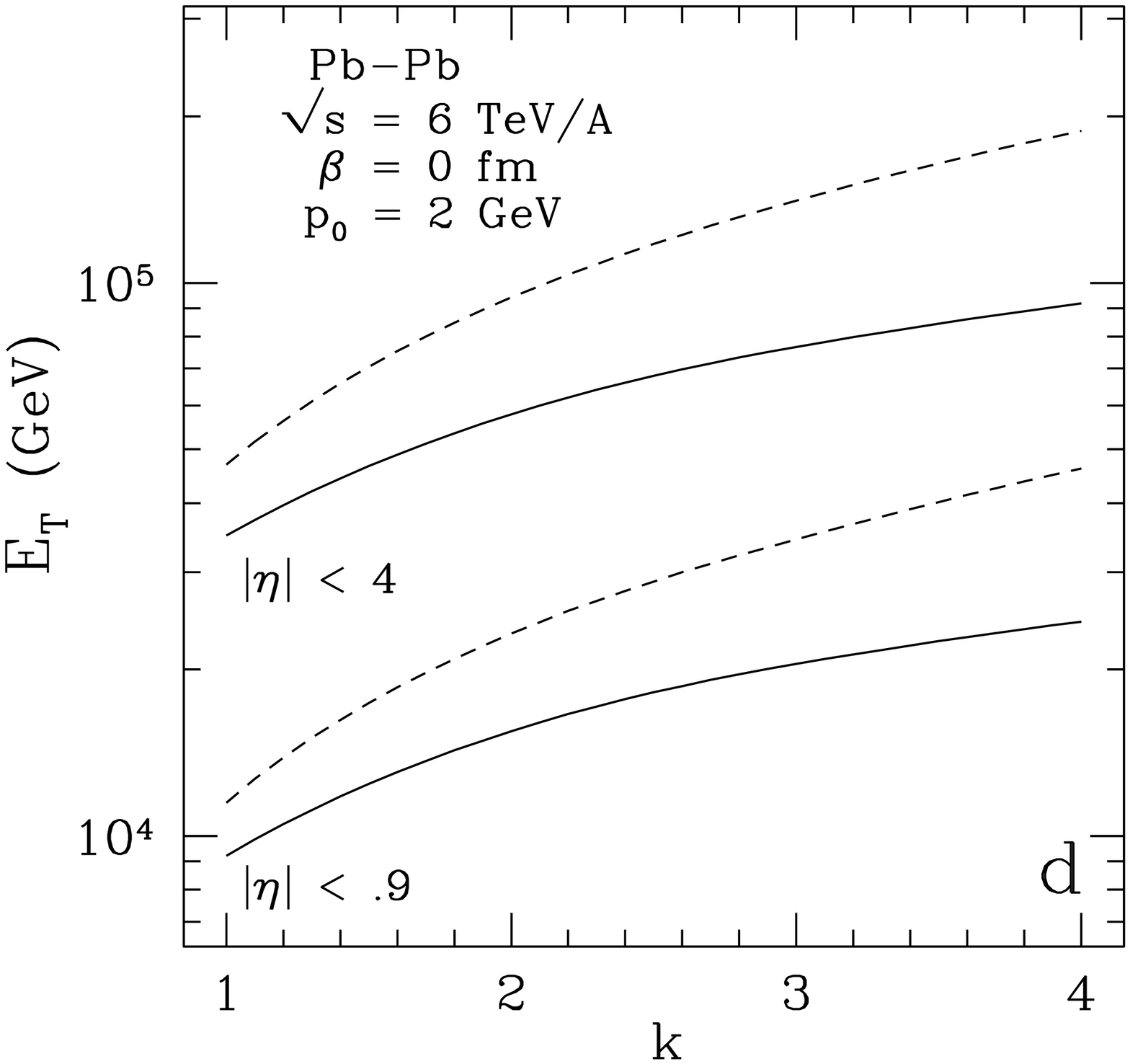,height=2.5in}
\caption{\footnotesize Average number $N$ of minijets, $a)$ $b)$,
 and average transverse energy $E_T$, $c)$
$d)$, in a Pb-Pb central collision in two different rapidity
windows, $|y|<4$ and $|y|<.9$, as a function of the cutoff, $a)$
and $c)$, and of the $k$-factor, $b$) and $d)$. The dashed curves
are computed without including rescatterings, the continuous curves
after including rescatterings. \label{totals}  }
\end{center}
\end{figure}

The effect of semi-hard parton rescatterings on the
average number of minijets and on the average transverse energy
produced in a central Pb-Pb collision at LHC energies is
summarized in Fig.\ref{totals}. We plotted the average number
of minijets and their transverse energy in the case of a very central rapidity
window, $|y|<.9$, corresponding to the ALICE detector, and in a
larger rapidity interval, $|y|<4$, that will be covered by the CMS
detector. Figure $\ref{totals}a)$ shows, in the two cases,
the dependence of the average number
of produced minijets on the choice of the cutoff $p_0$.
The dashed curves are the results obtained by the single scattering
expression, Eq.(\ref{pqcd}), while the continuous curves are the result
of the inclusion of semi-hard parton rescatterings, Eq.(\ref{W}) plus
the analogous term for the $B$-partons. These
expressions have been computed with the GRV98LO distribution
functions \cite{GRV98} with no shadowing corrections included,
and by representing the elementary partonic
interaction at the lowest order in QCD; to account for higher
order corrections the result of the elementary interaction has
been multiplied by a factor $k=2.5$. Both the $k$-factor and the scale
$Q=p_0/2$ where chosen in order to reproduce the value of the $p\bar{p}$
mini-jet cross-section at $\sqrt{s}=900$ GeV \cite{Albajar:1988tt}. The
sensitivity to the value of the $k$-factor is shown in Fig. $\ref{totals}b)$,
where the curves have the same meaning as in Fig. $\ref{totals}a)$, and the
cutoff has been fixed to the value $p_0=2$ GeV. Analogous curves for the
average transverse energy carried by the produced minijets are shown in
Fig. $\ref{totals}c)$ and $\ref{totals}d)$. The average transverse energy
without rescatterings has been computed by using Eq.(\ref{Et}) (dashed curves)
and with rescatterings (continuous curves) by using
Eq.(\ref{etr}), after integrating on $b$ and on $x$ (inside the
corresponding rapidity windows) and adding the analogous
contribution of the $B$-partons.


\section{Conclusions}
\setcounter{equation}{0}

\begin{figure}[tbh]
\begin{center}
\epsfig{figure=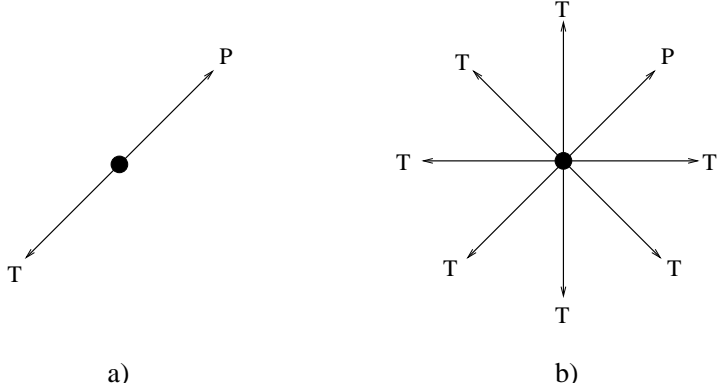,height=2in}
\caption{\footnotesize
Production of minijets in the transverse momentum plane. a) The projectile
parton (P) is allowed to interact only once with a target parton (T), so that
the two produced minijets are back-to-back.
b) The projectile can scatter against more than one target and minijets are
produced with a star-shape.
  \label{star-shape}  }
\end{center}
\end{figure}

In this paper we have discussed the production of minijets in
heavy ion collisions at the LHC, by including explicitly semi-hard
parton rescatterings in the dynamics of the interaction. The
regime of interest is, in fact, the regime where the nuclear
target reaches the black disk limit for a projectile parton
interacting with momentum exchange above the cutoff $p_0$ (which
defines the lower limit in $p_t$ for minijet production). In the
mechanism of production of minijets usually considered (a
projectile parton interacts with only a single target parton) the
elementary interaction generates a state with a preferred
direction in transverse space (the direction of the two minijets)
which is not consistent with local symmetry in the transverse
directions implied by the black disk limit. The star-like shape in
transverse space of the state produced by the multiple
interactions of a projectile with different target partons
(scattered projectile + recoils) recovers, on the other hand, the
symmetry property of the interaction in the black disk limit, see
Fig.\ref{star-shape}.

The basic element in our estimate is to recognize that the usual
expression that gives the average number, $\langle N(\beta)\rangle$, of parton
collisions at a fixed value of the nuclear impact paramete, Eq.(\ref{pqcd}),
is obtained by convoluting the average number of partons of the projectile
with the average number, $\vev{n_B(x,b)}$, of interactions of each projectile
parton with the target nucleus, Eq.(\ref{nb}). Notice that the
evaluation of $\vev{n_B(x,b)}$ and its dependence on the momentum
fraction is, in this way, determined in a unique way by the parton
distributions. Our results are therefore parameter free. Given the
expression of the semi-hard cross section, Eq.(\ref{csect}), the
average number of produced minijets, $W(x,b,\beta)$, Eq.(\ref{W}),
and the corresponding average transverse energy
$\vev{p_t(x,b,\beta)}$, (Eq.\ref{etr}), are computed without
further approximations, so that our result for $N$ and $E_T$,
plotted in the figures, are exact consequences of the nuclear
cross section (\ref{csect}). An approximation that is done when
writing the nuclear cross section (\ref{csect}), is to evaluate
the parton distributions at the scale of the cutoff, so that in
Eq.(\ref{dWdp}) the only dependence on the transverse momenta is
in the elementary partonic cross sections, which basically
corresponds to neglecting the logarithmic dependence of the
distributions in comparison with the inverse power dependence of
the cross section. However, our formalism can be extended to take
into account a general scale dependence.

The main features of the numerical evaluation are that semi-hard rescatterings
have a sizable effect on the average number of minijets and on the
transverse energy produced in heavy ion collisions at the LHC,
so that they affect also global
characteristics of the typical inelastic event. The induced
correction increases with the value of the $k$-factor, which
represents higher orders in the elementary parton collision, and
with the size of the rapidity window, since rescatterings are more
frequent for partons with a larger momentum fraction. By looking
at the dependence on the cutoff, both the average number of
minijets and the corresponding average transverse energy are more
regular at low $p_0$, showing a tendency to saturate below
$p_0 \simeq 3$ GeV and making in this way
the choice of the cutoff less critical.

\vskip.25in
{\bf Acknowledgment}
\vskip.15in
This work was partially supported by the Italian Ministry of University and of
Scientific and Technological Researches (MURST) by the Grant COFIN99.

\newpage


\end{document}